\documentclass{iaus}
\usepackage{graphicx}

\title[Light curve and orbital period analysis of AT Peg] %% give here short title %%
{Light Curve and orbital period analysis of the eclipsing binary AT Peg}

\author[Alexios Liakos, Panagiotis Niarchos and Edwin Budding]   %% give here short author list %%
{Alexios Liakos$^1$,
\ Panagiotis Niarchos$^1$
 \and Edwin Budding$^{2,3}$}

\affiliation{$^1$Department of Astrophysics, Astronomy and Mechanics, University of Athens, Athens, Hellas \\
email: {\tt alliakos@phys.uoa.gr,~}{\tt pniarcho@phys.uoa.gr } \\[\affilskip]
$^2$Carter Observatory \& School of Chemical and Physical Sciences, Victoria University, Wellington, New Zealand \\[\affilskip]
$^3$Department of Physics and Astronomy, University of Canterbury, Christchurch, New Zealand  \\
email: {\tt budding@xtra.co.nz}}

\pubyear{2011}
\volume{282}  %% insert here IAU Symposium No.
\pagerange{111--112}
\jname{From Interacting Binaries to Exoplanets: Essential Modeling Tools}
\editors{A.C. Editor, B.D. Editor \& C.E. Editor, eds.}
\begin{document}

\maketitle

\begin{abstract}
CCD photometric observations of the Algol-type eclipsing binary AT Peg have been obtained. The light curves are analyzed with modern techniques and new geometric and photometric elements are derived. A new orbital period analysis of the system, based on the most reliable timings of minima found in the literature, is presented and apparent period modulations are discussed with respect to the Light-Time effect (LITE) and secular changes in the system. The results of these analyses are compared and interpreted in order to obtain a coherent view of the system's behaviour.

\keywords{methods: data analysis, (stars:) binaries (including multiple): close, (stars:) binaries: eclipsing, stars: evolution, stars: individual (AT Peg).}
\end{abstract}

\firstsection % if your document starts with a section,
\section{Observations and analyses}

The system was observed during 6 nights in summer of 2010 at the Athens University Observatory, using a 20-cm Newtonian telescope equipped with the CCD camera ST-8XMEI and B and R Bessell photometric filters.

%\section{Light curve and O--C analyses}

The light curves (hereafter LCs) have been analysed with the PHOEBE v.0.29d software (\cite[Pr\v{s}a \& Zwitter 2005]{PZ05}). The temperature of the primary component was used as a fixed parameter (T$_1$=8400~K), based on the classification of \cite[Maxted et al. (1994)]{M94}. The spectroscopic mass ratio q of \cite[Maxted et al. (1994)]{M94} was used as initial value, and then it was adjusted, but always kept inside the spectroscopic error. The rest parameters were either given theoretical values or they were adjusted (for details see \cite[Liakos et al. 2011]{L11}). The contribution of a third light was also considered as there are indications of a third star orbiting the eclipsing pair. The absolute parameters of the components were calculated and used for further study of their present evolutionary status.

The least squares method with statistical weights in a MATLAB code (for details see \cite[Zasche et al. 2009]{Z09}) has been used for the analysis of the O--C diagram. The current O--C diagram of AT Peg includes 276 times of minima taken from the literature. A LITE fitting function, corresponding to a cyclic variation of the O--C points, and a parabola, assuming a mass-transferring configuration, were chosen to fit the times of minima.

\begin{figure}[t]
\begin{center}
\begin{tabular}{cc}
\includegraphics[width=5.3cm]{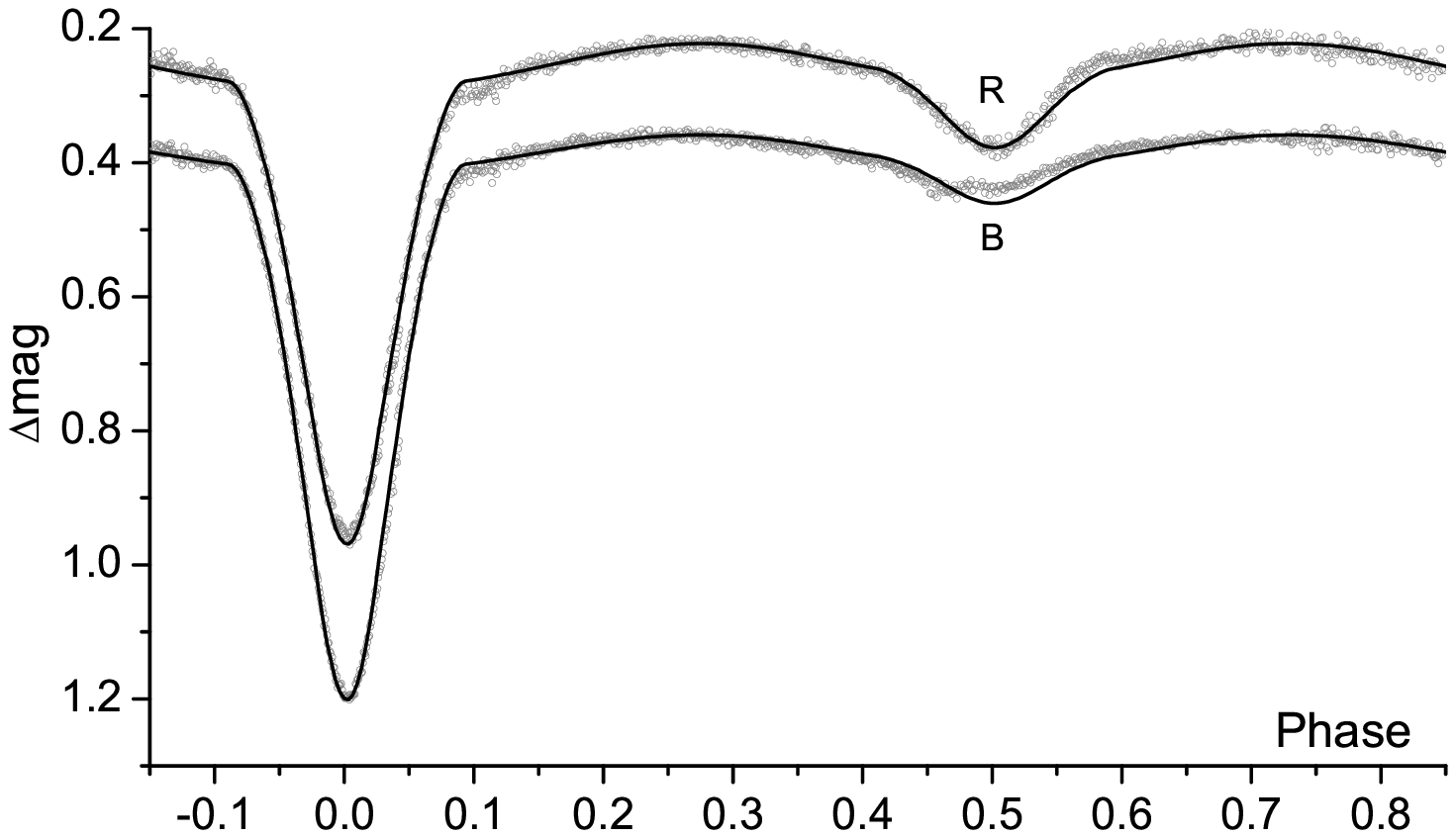}&\includegraphics[width=7.5cm]{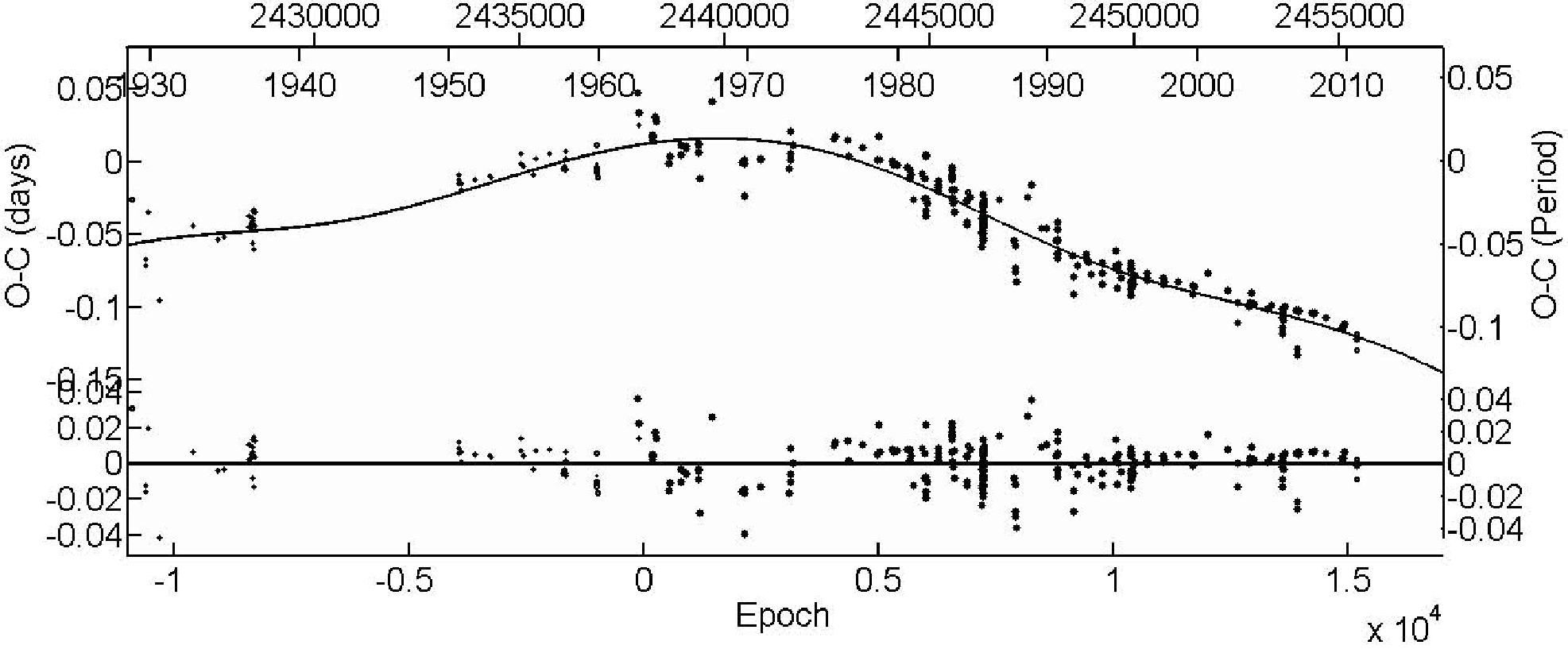}
\end{tabular}
\caption{Left panel: Observed (symbols) and synthetic (solid lines) light curves of AT Peg. Right panel: The O--C diagram of AT Peg fitted by the combined function (solid line).}
\label{fig1}
\end{center}
\end{figure}

The synthetic and observed LCs and the O--C fitting curve and its residuals are shown in Fig.\,\ref{fig1}, while the derived parameters from both analyses are listed in Table \,\ref{tab1}.

\begin{table}
\begin{center}
\caption{Results of light curve and O--C analyses and absolute parameters of the components.}
\label{tab1}
\scalebox{0.75}{
%{\scriptsize
\begin{tabular}{lcc lcclcc}
\hline
            \multicolumn{6}{c}{\bf Light curve parameters}     &\multicolumn{3}{c}{\bf Absolute parameters} \\
\hline
\emph{Component:}  & \emph{Primary}&\emph{Secondary}&\emph{Filters:} &\emph{B} & \emph{R}&\emph{Component:}  & \emph{Primary}&\emph{Secondary}   \\
\hline
T [K]          &   8400* &5189 (7)&     $x_{1}$    &  0.556  &  0.413  &M [M$_{\odot}$]& 2.2 (1)  &   1.0 (1)\\
g              &    1*   & 0.32*  &     $x_{2}$    &  0.834  &  0.596  &R [R$_{\odot}$]&  1.70 (3)&  2.14 (3) \\
A              &    1*   & 0.5*   & L$_1$/L$_T$    &0.809 (2)&0.727 (2)&L [L$_{\odot}$]&  13.0 (4)&  3.0 (1) \\
$\Omega$       &4.49 (1) & 2.83   & L$_2$/L$_T$    &0.115 (1)&0.212 (1)&M$_{bol}$ [mag]&  2.0 (2) & 3.6 (2)   \\
i [deg]        &\multicolumn{2}{c}{77.54 (5)}&L$_3$/L$_T$&0.076 (1)&0.062 (2)&a [R$_{\odot}$]& 2.18 (3)& 4.61 (9)  \\
q              &\multicolumn{2}{c}{0.478 (3)}&     &        &             &log$g$ [cm/s$^2$]&4.31 (3)&3.79 (3)    \\
\hline
                            \multicolumn{9}{c}{\bf O--C diagram parameters}                 \\
\hline
Min.I [HJD]       &         243803.447 (1)       &\.{P} [days/yr] &-3.563 (1) $\times 10^{-7}$ &$\omega$ [deg]& 204 (37) &f (M$_3$)[M$_{\odot}$]& 0.0129 (3)  \\
P [days]          &          1.1460905 (2)       &  P$_3$ [yrs]   &         49.7 (9)           &    A [days]  & 0.018 (1)& M$_{3,min}$ [M$_{\odot}$]& 0.57 (1)\\
C$_2$ [days/cycle]& -5.590 (1) $\times 10^{-10}$ &  T$_0$ [HJD]   &         2446308 (1796)     &        e     & 0.1 (1)   \\
\hline
\multicolumn{9}{l}{*assumed, L$_T$=L$_1$ + L$_2$ +L$_3$}
\end{tabular}}
\end{center}
\end{table}

\section{Discussion and conclusions}

New photometric and O-C diagram analyses of the eclipsing binary AT Peg were performed. The results were combined with those of previous spectroscopic studies and new geometric elements of the system and absolute parameters of its components were derived.
The primary component was found to be a dwarf star and close to the ZAMS limit. On the other hand the secondary is located beyond the TAMS line indicating that it is at the subgiant stage of evolution. Moreover, according to the LC analysis, this component was found to fill its Roche lobe, therefore it may be expected to be a mass loser. Hence, according to these results AT Peg can be considered as classical Algol type system.

The period changes analysis suggests two main conclusions: {\bf (a):} The LC solution revealed a third light of $\sim$7\%, while the O--C analysis suggested a third body with a minimal mass of 0.57 M$_{\odot}$. Assuming a MS nature with coplanar orbit and taking into account the Mass-Luminosity relation for dwarfs (L$\sim$M$^{3.5}$)  the expected light contribution was found $\sim$1\%, which is much less than the observed one. However, if the third body orbits the eclipsing binary with an inclination of $\sim$12$^{\circ}$ (as a MS star), or it is more evolved, then the observed additional luminosity can be justified.
{\bf (b):} The orbital period secular change found, contrary to what it was expected (the mass should flow from the less massive to the more massive component), resulted in a decreasing period rate. This discrepancy could be explained with a mass loss rate from the system of 6.6$\times$10$^{-7}$ M$_{\odot}$/yr, perhaps due to strong stellar winds, or with a magnetic breaking mechanism or even with systemic angular momentum loss, which superimposes the expected mass transfer.

%\begin{discussion}

%\end{discussion}

\end{document}